# The disruption index does not measure scientific innovation

Julien Larrègue[1], Yves Gingras[2]

## Introduction

A paper recently published in *Science* under the rubric of *"Policy Article"* argued that what they call "scientific disruption" declines with academic age, and that this decline is related to the absence of mandatory retirement for older academics (Cui et al., 2026). Since its publication, the article's conclusions and policy suggestions in relation to mandatory retirement have received considerable media attention. *Nature News* ran the headline "Early-career researchers do more 'disruptive' science than veterans" (Lenharo, 2026). *The Economist*, adding a geopolitical framing, concluded that "At a time of great-power rivalry, when countries are racing to the technological frontier, it is helpful to have youth on your side" (2026). These are just two examples. Several other news articles have covered this publication and commented on its findings (Grant, 2026; Oza, 2026; Singh Chawla, 2026).

Thus, it is worth taking a closer look at the proposed measure of "disruption" since all the analysis and conclusions are based on the results obtained from this index, thus taking it as valid. The issues we address are not specific to this article and can be found in many papers using bibliometric data that propose a new "index" on the basis of common sense intuition and then using it as a "black boxed" instrument to measure "quality," "innovation" or, now, "disruption" for creating "rankings" and formulate policy actions on the basis of the calculated values of the index. Among the best known cases are of course the "h-index", much criticized by bibliometricians but nonetheless constantly used by scientists and managers, or the Shangai ranking with its many problematic indicators supposed to rank the best universities in the world (Gingras 2016).

We thus propose here to open the black box of the "disruption index" to see if what it is supposed to measure corresponds to the concept of "disruption" behind the index. As the sociologist Paul Lazarsfeld (1958:101) underlined decades ago, "the flow of thought and analysis and work which ends up with a measuring instrument usually begins with something which might be called imagery," but that vague intuition must be analyzed carefully to make sure the chosen index provides a valid measure of the concept it is supposed to represent. For instance, if one wants to measure temperature, one much construct a thermometer; but if one wants to measure humidity, one needs an hygrometer. So, before taking an index as valid and use it, one must first make sure that its properties are consistent with the content and meaning of the concept one wants to empirically measure (Gingras, 2016). The disruptive index being based on citation practices, let us first recall what we know of this practice.

[1] Département de sociologie, Université Laval – Centre interuniversitaire de recherche sur la science et la technologie (CIRST), julien.larregue@ulaval.ca
[2] Centre interuniversitaire de recherche sur la science et la technologie (CIRST), UQAM, gingras.yves@uqam.ca.

Sociologists of science and bibliometricians have long emphasized that the act of citing is a complex social practice and, as Nigel Gilbert (1977: 114) observed fifty years ago, we do not have "an explicit and accepted theory concerning the reasons why scientists normally cite other papers, and why authors choose to cite particular papers rather than others. Accordingly, we do not yet have any clear idea about what exactly we are measuring when we analyze citation data." The many studies of citation practices that have been published since that observation confirmed that the diversity of reasons to cite (or not) a paper cannot be summarized in a single indicator that could then be interpreted as simply as temperature on a thermometer (Cronin, 1984; Gingras et al., 2009; Larivière et al., 2009; Lozano et al., 2012; McCain, 2014). Thus, without a clear understanding of why researchers cite (or not) some articles more than others, measures derived from citation patterns remain uninterpretable and/or based on unvalidated a priori beliefs. This critique applies to all of the disruption indices that have been used in the literature in recent years (for a recent review, see Leibel and Bornmann, 2024). We argue that such indices should not be taken for granted or prematurely "black-boxed" as valid measures of scientific innovation or paradigm-shifting research and that, as a concequence, they should not be the basis of policy decisions.

Given the large media circulation of the (often implicit) conclusions suggested by the results presented in Cui et al.'s article, we will look in detail at its content. Our analysis reveals theoretical, conceptual and methodological issues, implausible results, as well as selective reporting of analyses in support of the authors' narrative. Our conclusion is that the authors' claims and the policy implications they formulate do not really follow from their data. We focus here on four key elements that raise serious doubts about the claims put forth in this article. All of these issues stem from the failure to treat citation as a complex social practice : 1) the conceptual meaning and methodological construction of the disruption index itself, 2) the claimed relationship between the age of scientists and the proportion of disruptive articles across countries, 3) the high disruptive rate exhibited among early-career scholars across disciplines, and 4) the purported connection between academic retirement policies and scientific disruption. In our concluding remarks, we address the lack of conceptual reflection in the use of complicated scientometric measures, a weakness that is frequently overshadowed by unnecessarily technical debates that take unverifiable and implicit assumptions for granted.

### 1) What does “disruptive” mean?

Following previous literature, the authors define disruption as “the replacement of established ideas with new ones” (Cui et al., 2026: 588). They distinguish it from novelty, which consists in “the linking of previously unconnected ideas” (Cui et al., 2026: 588). We focus here on disruption alone, since the authors center their arguments around this notion. Two intertwined issues are worth discussing: one is conceptual, the other is methodological.

The first issue is that the definition of disruption remains vague throughout the article. Instead of providing conceptual clarity, the article relies on a formal network-based citation measure that distinguishes between neutral, developmental and disruptive articles according to citation structures:

> “The Disruption (D) index is a citation network-based measure identifying breakthrough papers (Funk and Owen-Smith, 2017; Wu et al., 2019). The underlying intuition is to quantify how subsequent papers cite a focal paper while disregarding its references. The D-index is calculated by categorizing subsequent papers citing a focal paper $f$ into three types: type $i$ papers that only cite the focal paper, ignoring its references; type $j$ papers that cite both the focal paper and its references; and type $k$ papers that only cite the references, excluding the focal paper.” (Cui et al., 2026, supplementary materials, p. 6)

This is schematically represented in Figure 1, with $D$ = 1 being indicative of a disruptive article.

**Figure 1. Representation of the disruption index**

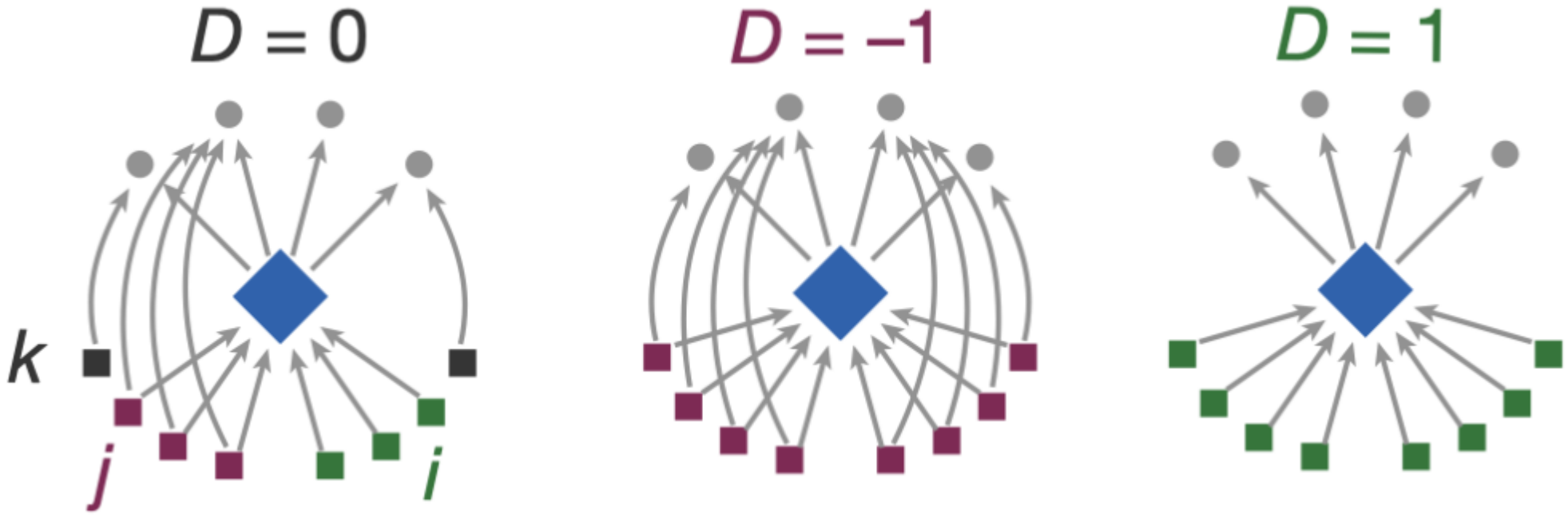

Source: (Wu et al., 2019: 379)

Thus, the author's categorization of (non)-disruptive articles rests more on citation practices than on conceptual analysis of articles' content. Since they provide no explicit theorization of citation practices to justify their categorization of articles, we are led to wonder what the index really measures. Why should we assume that citing an article instead of its references is evidence that the article is more original or groundbreaking than the prior works it draws upon? As Cronin (1984: 28–29) noted four decades ago, "The essential subjectivity of the act of citing means that the reasons why an author cites as he does must remain a matter for conjecture." The risk here is to reverse the order of thinking by black-boxing the notion of "disruption" and taking it for granted before having first seriously thought about the intellectual and conceptual meaning of "innovation" and "disruption" – which suggest a radical innovation and a certain level of intentionality on the part of the revolutionary scientist.

Hence, on strictly logical grounds, it is impossible to infer – like the authors do – that a value of $D$ = 1 is indicative of scientific disruption instead of mere bibliographic changes. Such an interpretation does not follow from the citation data. It rests on an additional assumption about the meaning of citation behavior that remains to be demonstrated. This is especially problematic insofar as past research has showed that scientific breakthroughs are not a homogenous category. In fact, some findings indicate that "most scientific breakthroughs are driven by an already existing question and in line with the state-of-the-art literature," which makes them belong to Kuhnian "normal science" (Wuestman et al., 2020: 1216).

A closer look at the construction of the index confirms that this measure does not capture scientific innovation. We gather from the supplementary materials that the authors used the version of the D-score found in Wu et al. (2019), a measure usually referred to as $D_1$. An article has a positive D-index whenever $N_i > N_j$, that is when more future papers cite the focal article alone instead of citing it alongside the original references it cited. It does not require that the focal article be highly cited or influential in any meaningful sense. An article can have a positive disruption score even if only three future citing articles are involved:

$$D = \frac{2-1}{3} = 0.33$$

Of course, no one would consider this focal article to be a major scientific breakthrough. This contradicts the authors' claim that the disruption index they use makes it possible "to operationalize Kuhnian paradigm shift" (Cui et al., 2026, supplementary materials, p. 7). In fact, it simply characterizes a bibliographic shift, for which there can be many explanations. One cannot expect that using millions of papers and making averages will correct for all these variations and lead to a well defined measure capturing originality and "disruption". Other alternate measures of disruption were developed in the past few years to account for the limitations of $D_1$ (see for instance Deng and Zeng, 2023). A recent review article on the many proposed citation-based measures of scientific disruption reached the conclusion that "the $D_1$ and its variants have some considerable limitations that they share with many other bibliometric indices" (Leibel and Bornmann, 2024: 632). This measure is affected by several factors, including time of publication, discipline, document type, and language. More importantly, it takes for granted that citing is a measure of quality and innovation, thus ignoring the persuasive and symbolic dimension of referencing (Gilbert, 1977; Small, 1978). As we shall see now, the country-level findings make it highly improbable that the disruption index captures paradigm shifting research.

### 2) Are Iraq and Nigeria the most "disruptive" countries in science?

The issues raised by the $D_1$ index and the conceptual ambiguity of the notion of "disruption" (whose connotation suggests a strong innovation) become even more obvious when we look at the results presented for different countries. According to the authors, "Countries with younger scientific workforces (e.g., China, India) tend to produce more disruptive work, whereas those with older workforces (e.g., the United States, United Kingdom) tend to integrate and recombine knowledge" (Cui et al., 2026: 591). This conclusion rests on a figure reproduced below (Figure 2), where we indeed observe a negative relationship between the average career age of authors and the percentage of disruptive articles across countries.

**Figure 2. Correlation between percentage of disruptive articles and average career age of authors across countries**

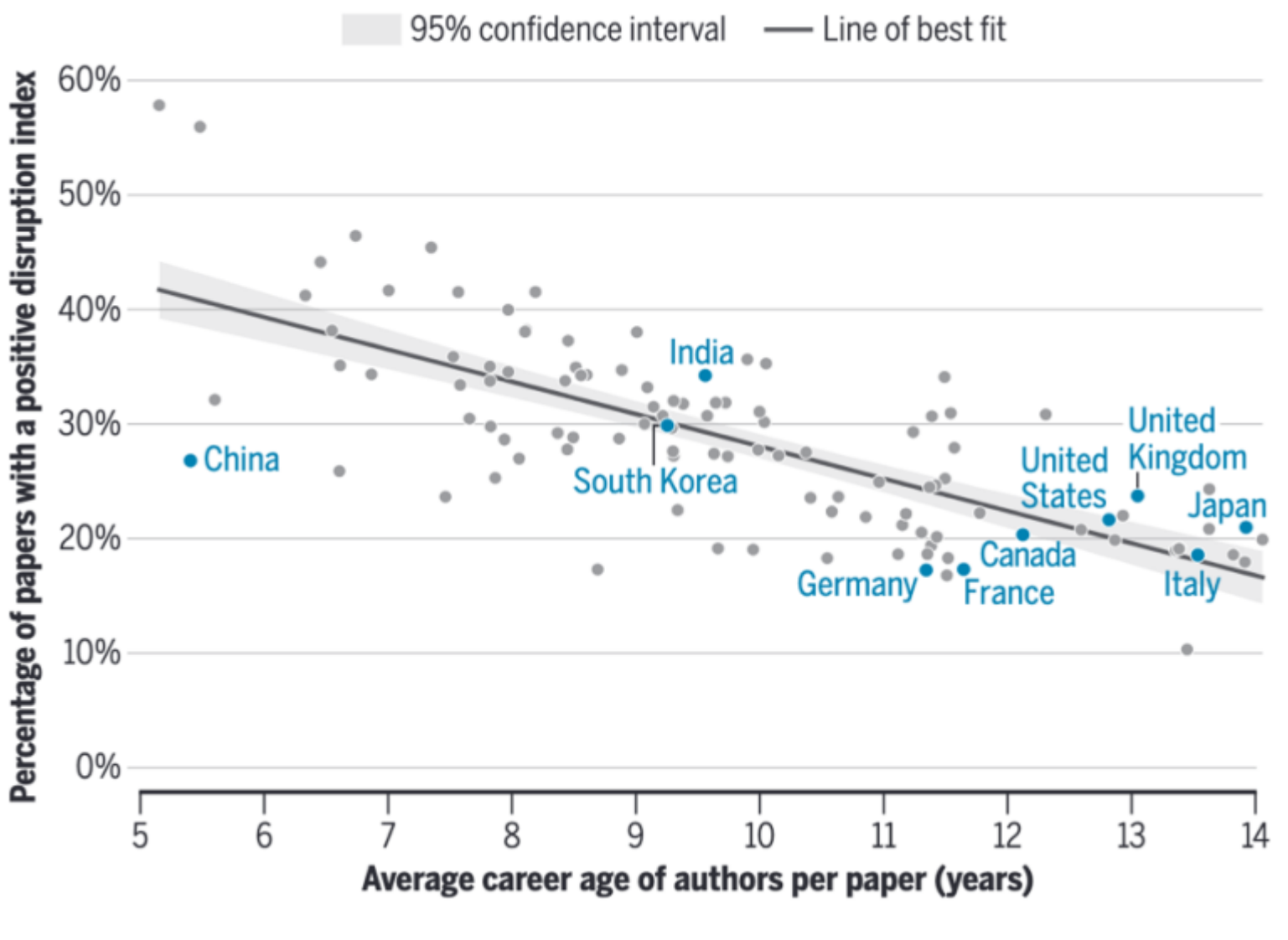

Source: (Cui et al., 2026: 590)

Curiously, none of the most efficient countries in terms of disruption (top left above the regression line) are labelled or discussed in the paper. We thus reproduced the figure from the authors' own data[3] and added missing labels (Figure 3). As we can see, the most disruptive countries in science are Iraq and Nigeria, with 57.5% and 56% of articles exhibiting a positive disruption index, respectively. India and China, discussed by the authors are far from reaching the same levels of disruption. They rank respectively 26th and 66th, with 34.3% and 26.9% of articles exhibiting a positive disruption index. It is unclear what accounts for the substantial gap between Iraq and Nigeria on the one hand, and China on the other, given the similar average academic ages of their research workforces (5.2,

[3] https://osf.io/8m7pg/overview.

5.5, and 5.4 years, respectively). We have here a first indication that the behavior of the index does not match what we could expect from a “disruptive” article – one that is very innovative, based on up to date and high quality resources and instruments. Of course, this is something that a simple bibliographic change cannot detect.

**Figure 3. Correlation between percentage of disruptive articles and average career age of authors across countries (2010)[4]**

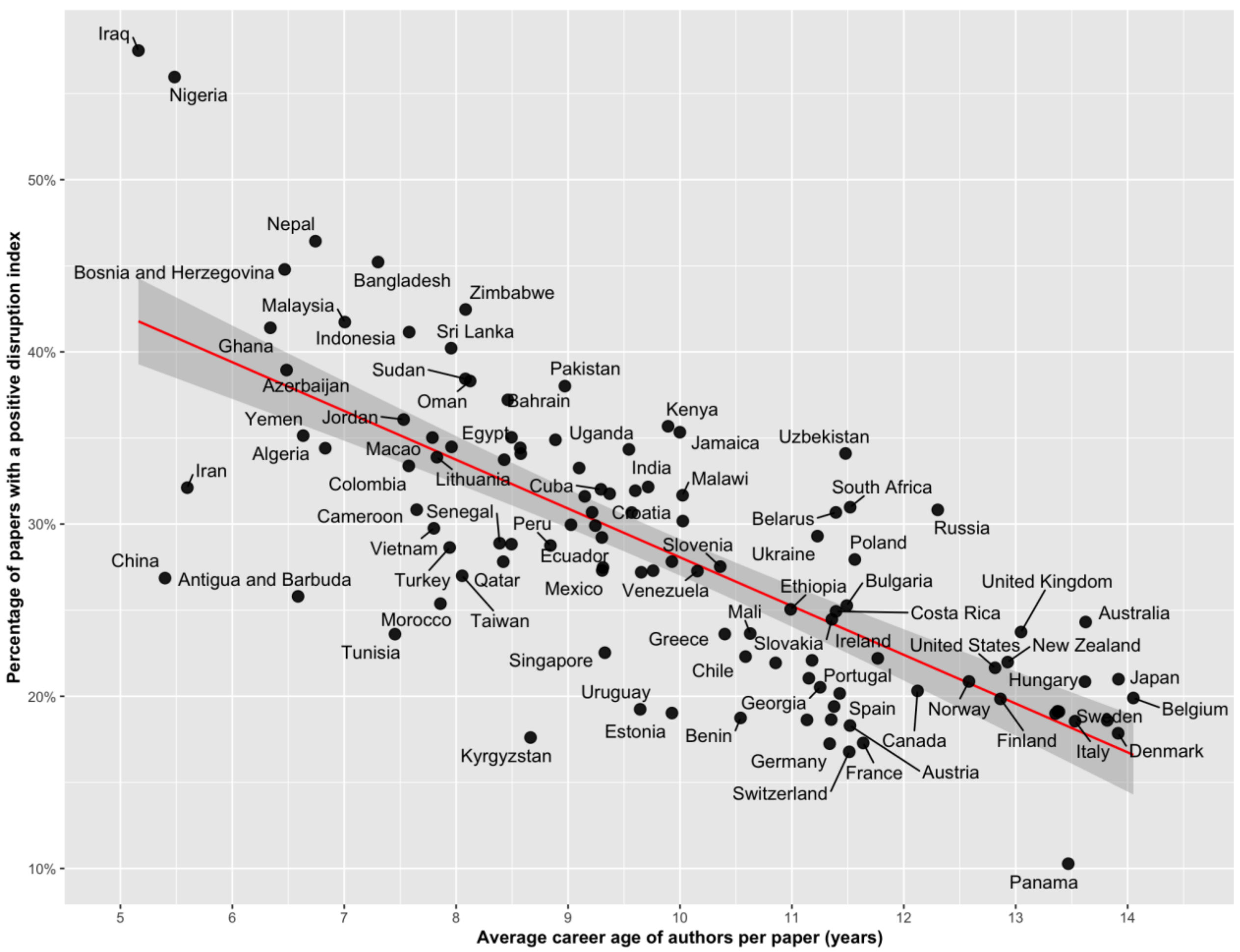

These findings can be interpreted in at least three different ways:

*Hypothesis 1. Iraqi and Nigerian scientists are producing scientific breakthroughs at very high rates.*

[4] Following the authors’ design, only countries with 200 authors and more are included.

*Hypothesis 2. The D-index captures complex referencing practices that are peripheral and even external to the scientific and conceptual content of the paper and thus unconnected to any putative scientific breakthrough.*

*Hypothesis 3. The D-index captures the replacement of internationally shared references with local references.*

Since the authors do not discuss Iraqi and Nigerian science, we assume that they do not believe *Hypothesis 1* to be pertinent. We also note the absence of the names of many countries that happen to be much more “disruptive” than the few labelled OECD countries (France, Germany, Canada, etc). In so doing, the authors implicitly acknowledge that the D-index does not really measure conceptual and paradigm shifting research. Otherwise, why would they contrast USA with China and India, two countries with comparatively much lower disruption rates? For if the index is valid, there is of course no reason to believe that Iraq and Nigeria are not more efficient at publishing disruptive papers than the United States. If one objects that the index is not valid for these countries, then that would confirm that it is not a valid measuring instrument, and that it is a priori beliefs that lead one to decide when the indicator is worthwhile.

The findings do seem to indicate that another dynamic is a play. While Iraq and Nigeria score particularly high on the positive disruption index, a broader look at the distribution reveals that the left side of the curve – that is, the region containing the highest concentration of disruptive articles – is populated primarily by countries with relatively small and/or underfunded scientific systems. This pattern is evident when examining national publication and authorship data. According to our analysis of the authors’ data, the share of disruptive articles is negatively correlated with both the number of authors and the number of publications ($r$ = -0.20 and $r$ = -0.21, respectively). The ten *most* disruptive countries average just 2,608 authors and 1,836 publications, compared with 40,348 authors and 27,196 publications among the ten *least* disruptive countries.

Far from reflecting exceptional scientific creativity, high disruption scores may therefore be partly associated with the structural characteristics of small and/or underfunded research systems that occupy peripheral positions within the scientific field (*Hypothesis 2*). Another possible explanation for this result is that the disruption index captures the replacement of widely cited references – including from dominant, Western

countries – with national references that are better aligned with local or regional concerns (*Hypothesis 3*). Because the concept of disruption is poorly specified, it is impossible to draw definite conclusions from these observations. In any case, the index provides no indication about the quality and conceptual innovation of the paper, a simple fact that is hidden behind an apparently complex equation relating cited and uncited references.

### 3) The disciplinary diversity of scientific innovation

The above analysis demonstrates that the disruption index cannot be considered a valid measure of countries' innovation capacity. Now we will show that the same problems emerge when we look at how disciplines vary in their propension to produce "disruptive" science. The authors focused on seven broad domains: biology, chemistry, computer science, mathematics, medicine, physics, social science. Their main conclusion is that "Among 64,385 scientists with careers spanning 40 years or more, the probability of producing a top 10% disruptive paper decreases sharply with career age across all fields, indicating that older scientists are less likely to overturn established ideas" (Cui et al., 2026: 589). But given that the disruption index does not really capture any eventual overturning of established ideas, all we can say on the basis of these data is that younger researchers tend to have different referencing practices than their senior colleagues. What this means empirically is however unclear, and any definite conclusion is nothing more than speculation, even though it may please readers who spontanoeuly accept the popular, common sense view that younger people are intrinsically more creative.

What we would like to highlight now is that the article proceeds as if all the sciences were the same. Although some fields are distinguished, the method and analysis stem from the implicit idea that both citations and paradigm shifts obey identical norms across research domains. Hence, the fact that about 25% of early-career researchers produce articles situated in the top 10% of disruptive papers is assumed to be indicative of a shared pattern across social science, mathematics, or physics. Another implicit – though problematic – assumption is that "disruption" means the same in all these very different fields: a breakthrough in sociology thus would look the same as one in physics, and we can capture both by detecting identical referencing patterns. The problem is that we know this assumption is empirically false. Breakthroughs in social science look different from breakthroughs in physics, which also can be distinguished from breakthroughs in computer science or medicine, all disciplines with very different instruments and evidence standards. In fact, even social sciences are far from being a homogeneous field. Recent studies of funding competitions across national contexts show that different understandings of "quality" and "originality" coexist and conflict with each other (Larregue and Nielsen, 2024; Larregue and Pavie, 2026).

The case of sociology is revealing because it brings into view another assumption underlying the authors' argument. Implicit in their reasoning is the notion that citing more recent publications is inherently preferable because it would suggest "innovation," which is supposed to reside at the edge of knowledge and therefore in recent and new papers. Yet this is not an empirical finding derived from the data; it is a normative judgment about what constitutes "good" research. As is well-known, some of the most fundamental works in sociology have been published several decades ago and are still widely cited. Émile Durkheim and Karl Marx, for instance, both born and active in the 19th century, remain to this day among the most cited sociologists globally (Gingras and Khelfaoui, 2025: 45–46). Whether this impedes scientific "disruption" is open to question, but so too is the assumption that "disruption" is necessarily based on recent papers displacing older ones.

### 4) Mandatory retirement does not cause innovation

To further explore the connection between academic age and disruptive ideas, the authors provide a comparative analysis between the United States and the United Kingdom. They resort to fixed-effects regressions to measure the change in the age of cited references caused by the end of mandatory retirement in the United States in 1994. They conclude that the "results rule out placebo effects and demonstrate that the 1994 end of mandatory retirement *caused* a sustained shift toward citing older literature among U.S. scientists, suggesting that longer academic careers of senior researchers contribute to slower renewal of canonical knowledge." (Cui et al., 2026, supplementary materials, p. 10). We will come back to the limitations of this US-centric narrative. But first, it is worth reviewing the results of this comparison limited to two countries.

As we can see in Figure 4 reproduced below, it is only for the years 1998 and 1999 that regression coefficients are statistically significant. Estimates for these years are also small, with a relative difference of about 0.3 years. It is unclear what the implications of this finding are supposed to be: does it indicate a reduction in paradigm-shifting research in the United States simply because, on average, cited articles are about *three months older* than those in the United Kingdom? The question is worth asking since the authors do not provide any direct measure of the age of researchers but only a proxy based on the time since their first publications. The caption of the figure reads: "Reduced turnover among senior scholars increases reliance on older literature" (Cui et al., 2026: 591). Put otherwise, they assume that the end of mandatory retirement automatically led to an aging of the research population in the United States.

**Figure 4. Evolution of the age difference between US and UK references**

Estimate | 95% confidence interval
Change in US-UK reference-age gap relative to 1994
0.4 years
0.2
0
−0.2
−0.4
1989
1991
1993
1995
1997
1999
Year

Source: (Cui et al., 2026: 591)

The problem with this affirmation is that it is well known that the average age of references in scientific papers has been *increasing* over the last century, and not *decreasing* as some scientists incorrectly believe (Larivière et al., 2008). This long term rise is of course not visible in the authors's analysis since it is restricted to a ten-year window, raising the question of whether the observed pattern persists, as we should expect, over a longer period for these two countries. Using data from the *Web of Science*, we reproduced the figure and extended it to 2024 for natural science, engineering, and medicine (Figure 5A). We also used the authors' *Open Alex* data to examine the evolution of the average age of references cited in articles published by US and UK researchers between 1974 and 2010 (Figure 5B). The results are strikingly similar and confirm the general trend mentioned above, which is also valid at the world level.

**Figure 5. Evolution of the average age of references cited in articles published by UK and US researchers**

*A. Web of Science data (Natural science, engineering and medicine)*

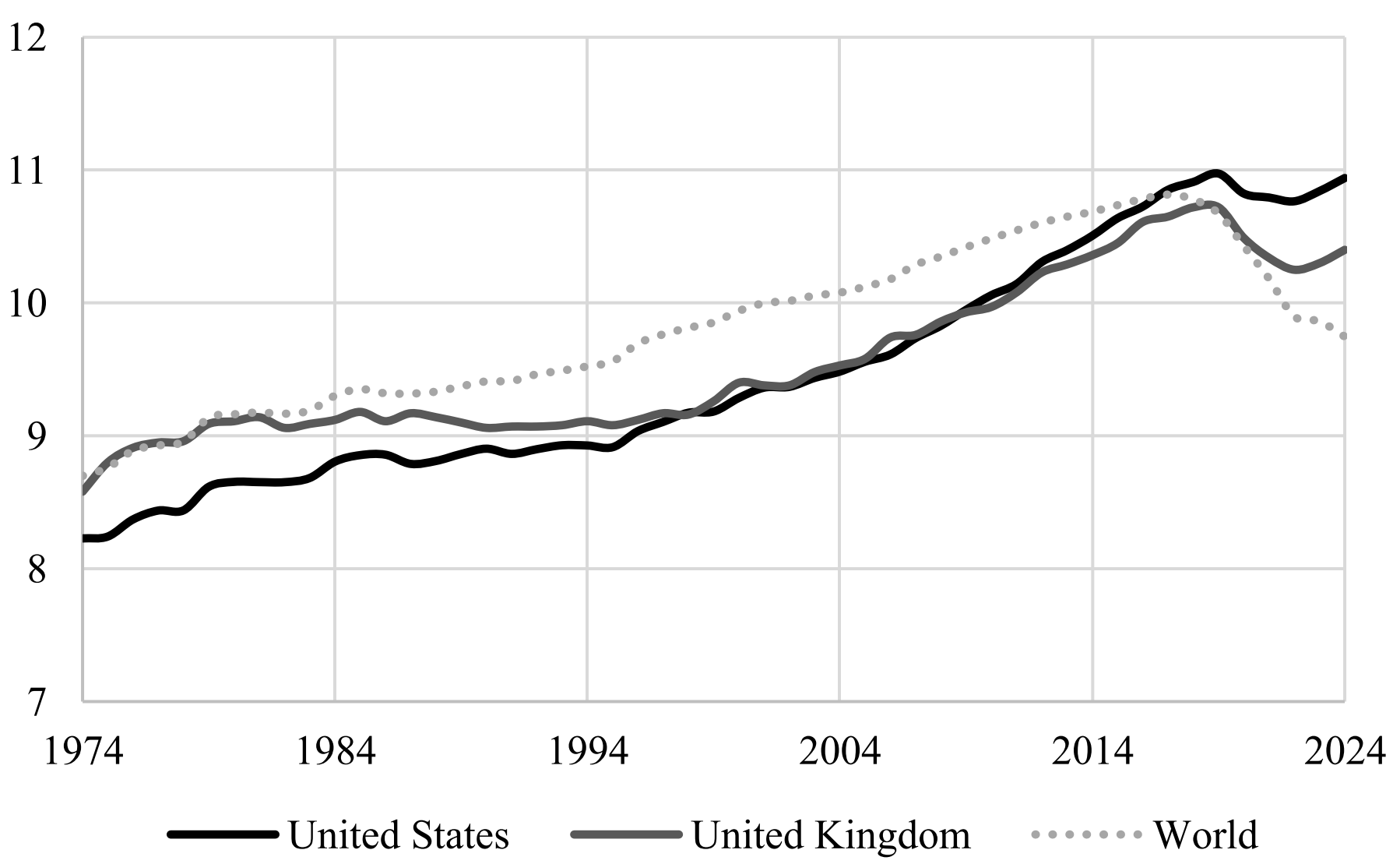

*B. Authors' OpenAlex data*

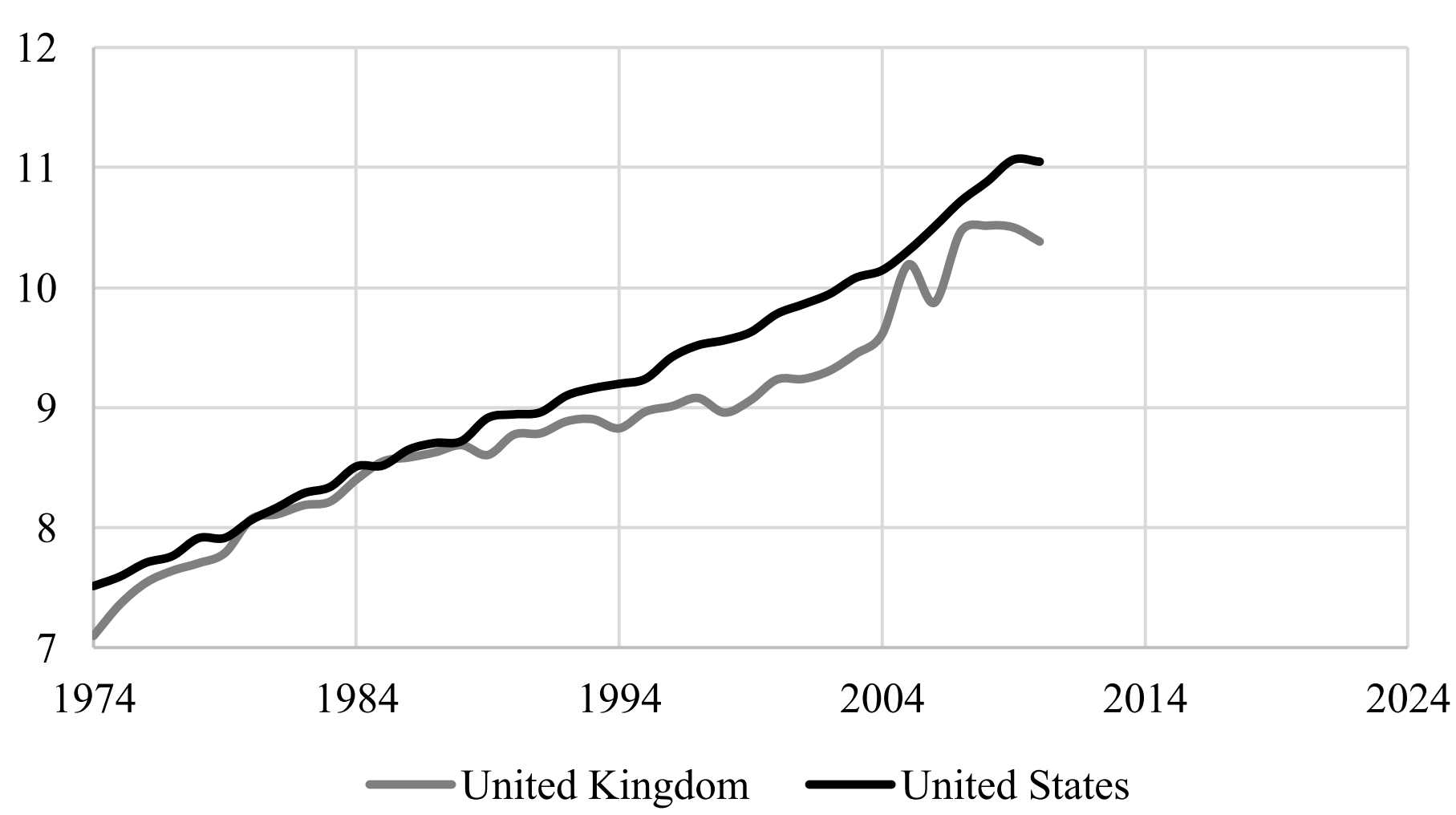

The comparison between the *Web of Science* and the authors' *OpenAlex* data show some minor differences that are likely due to the lower quality and reliability of the latter. In particular, unlike the *Web of Science*, *OpenAlex* does not include references that are not part of its database, which artificially lowers the average age of cited references in earlier periods. Despite these marginal differences, there is no doubt that the average age of cited references has increased steadily in both countries, with the two series exhibiting very high correlations over the full period in the *OpenAlex* dataset ($r = 0.97$). The correlation stays the same when we restrict the analysis to the 1995–2010 period alone ($r = 0.96$), that is after the 1994 US Supreme court decision ended mandatory retirement.

Importantly, this longer historical perspective also shows that the rise in reference age clearly predates the 1994 abolition of mandatory retirement in the United States. There is no visible break in trend or abrupt increase after 1994, and the modest divergence between the United States and the United Kingdom in the *OpenAlex* dataset begins around 1989. No difference at all is detectable in the *Web of Science* data, where the average age of cited references is almost identical for the 1995-2024 period (9.9 years for the United Kingdom, 10 years for the United States). In short, available bibliometric data do not support their claim that "the 1994 end of mandatory retirement *caused* a sustained shift toward citing older literature among U.S. scientists." (Cui et al., 2026, supplementary materials, p. 10).

Yet, for the authors, the policy implications of their findings are clear. In a *Nature* news article, they argued that "retirement norms in academia" should be reconsidered (Lenharo, 2026: 656). While they do not specify what measures should be adopted, we can infer from their article and public interventions that they would be favorable to returning to the US pre-1994 situation when retirement was mandatory. This US-centric narrative, as if the laws of the United States were universal and synonymous with "norms in academia," is problematic. Even a cursory international comparison casts doubt on the authors' reading. Several countries, including France, Germany, and Switzerland, maintain mandatory retirement policies for university professors, yet none exhibits higher levels of scientific disruption than the United States (see Figure 3).

Moreover, even supposing that countries with low disruption are often those without mandatory retirement, this would not imply – in good logic – that having mandatory retirement would mechanically increase disruption. If mandatory retirement were a key driver of disruptive research, one would expect these countries to outperform the United States on this measure. The fact that they do not, and the lack of clear effect of US legal reforms on the age of cited references, suggest that the authors' conclusions are more a priori convictions disguised as scientific calculations than logical inferences from their findings.

## 5) Concluding remarks

As we have argued throughout this paper, the "disruption" index cannot be considered a valid instrument since its meaning remain vague and, more importantly, it does not really measure what it is supposed to measure. As a consequence, it should not be used to infer any policy regarding innovation in science or the composition of the workforce. This is true not just of the article we have analysed here in detail, but of other versions of the index that assume that scientific breakthroughs can be deducted from citation and referencing patterns – whatever these may be. In these concluding remarks, we would like to stress out a more general tendency, in the field of scientometrics and so-called "science of science," to focus on technical issues at the expense of conceptual ones.

In the past few years, the disruption index as become increasingly used and critical articles have begun to appear as well. Some argued that the proposed index was biased by citation inflation and its time-sensitivity (Petersen et al., 2024); others noted that these indices did not correlate with researchers' self-assessments of their own work (Leibel et al., 2026), or that dataset artifacts led to erroneous findings (Holst et al., 2024). While these measurement issues are worth exploring and important, they tend to sideline a more fundamental issue: the notion of disruption itself is vague and prone to subjective assumptions that are then projected on it without regard to the actual content of the index. This reflex may explain the media reception of the paper, as its conclusions seem to confirm popular beliefs about age and creativity. And it is well known that when the conclusions of an article aligns with one's beliefs, one does not try to open the "black box" of the measuring instrument to make sure that it is indeed measuring what it is supposed to measure.

Thus, what is urgently needed is not another new or "better" measure, based on the uncritical use of citation or reference patterns, but a theoretical reflection on the meaning of the concept one wants to measure – here scientific change, innovation, originality, or what have you. Only then one should try to find a measure that is really *adequate* to the well defined concept. Otherwise, one risks behaving like a naïve scientist who tries to develop a more sophisticated hygrometer in the hope that it will provide a better measure of temperature.

**References**

Cronin B (1984) *The Citation Process: The Role and Significance of Citations in Scientific Communication*. London: Taylor Graham.

Cui H, Lin Y, Wu L, et al. (2026) Aging and the narrowing of scientific innovation. *Science* 392(6798): 588–591.

Deng N and Zeng A (2023) Enhancing the robustness of the disruption metric against noise. *Scientometrics* 128(4): 2419–2428.

Gilbert GN (1977) Referencing as Persuasion. *Social Studies of Science* 7(1): 113–122.

Gingras Y (2016) *Bibliometrics and Research Evaluation: Uses and Abuses*. Cambridge: The MIT Press.

Gingras Y and Khelfaoui M (2025) The relative (in)visibility of sociologists in the French, American, British, and German national fields (1970–2018). *Social Science Information* 64(1): 32–59.

Gingras Y, Larivière V and Archambault É (2009) Literature citations in the internet era. *Science* 323(5910): 36.

Grant B (2026) Is This Why Science Advances One Funeral at a Time? *Nautilus*, 11 May.

Holst V, Algaba A, Tori F, et al. (2024) Dataset Artefacts are the Hidden Drivers of the Declining Disruptiveness in Science. arXiv. Available at: https://arxiv.org/abs/2402.14583 (accessed 2 June 2026).

Larivière V, Archambault É and Gingras Y (2008) Long-term variations in the aging of scientific literature: From exponential growth to steady-state science (1900-2004). *Journal of the American Society for Information Science and Technology* 59(2): 288–296.

Larivière V, Gingras Y and Archambault É (2009) The decline in the concentration of citations, 1900–2007. *Journal of the American Society for Information Science and Technology* 60(4): 858–862.

Larregue J and Nielsen MW (2024) Knowledge Hierarchies and Gender Disparities in Social Science Funding. *Sociology* 58(1). SAGE Publications Ltd: 45–65.

Larregue J and Pavie A (2026) Prestige at Play: University Hierarchies and the Reproduction of Funding Inequalities. *Canadian Review of Sociology/Revue canadienne de sociologie* 63(1): e70012.

Lazarsfeld PF (1958) Evidence and Inference in Social Research. *Daedalus* 87(4). The MIT Press: 99–130.

Leibel C and Bornmann L (2024) What do we know about the disruption index in scientometrics? An overview of the literature. *Scientometrics* 129(1): 601–639.

Leibel C, Tekles A, Aksnes DW, et al. (2026) Is groundbreaking research disruptive? The covariation of disruption indices with researchers' self-assessments. *Quantitative Science Studies*: 1–22.

Lenharo M (2026) Early-career researchers do more 'disruptive' science than veterans. *Nature* 653(8115): 655–656.

Lozano GA, Larivière V and Gingras Y (2012) The weakening relationship between the impact factor and papers' citations in the digital age. *Journal of the American Society for Information Science and Technology* 63(11): 2140–2145.

McCain KW (2014) Obliteration by incorporation. In: Cronin B and Sugimoto CR (eds) *Beyond Bibliometrics: Harnessing Multidimensional Indicators of Scholarly Impact*. Cambridge: MIT Press, pp. 129–149.

Oza A (2026) Science is becoming less disruptive. Is an aging workforce to blame? *STAT*, 7 May. Available at: https://www.statnews.com/2026/05/07/young-researchers-more-innovative-new-study-says/ (accessed 4 June 2026).

Petersen AM, Arroyave F and Pammolli F (2024) The disruption index is biased by citation inflation. *Quantitative Science Studies* 5(4): 936–953.

Singh Chawla D (2026) Scientists produce more novel but less disruptive work as they age. *Chemistry World*, 2 June. Available at: https://www.chemistryworld.com/news/scientists-produce-more-novel-but-less-disruptive-work-as-they-age/4023585.article (accessed 4 June 2026).

Small HG (1978) Cited Documents as Concept Symbols. *Social Studies of Science* 8(3). 3: 327–340.

*The Economist* (2026) Why science is becoming less innovative. May. Available at: https://www.economist.com/graphic-detail/2026/05/25/why-science-is-becoming-less-innovative (accessed 1 June 2026).

Wu L, Wang D and Evans JA (2019) Large teams develop and small teams disrupt science and technology. *Nature* 566(7744): 378–382.

Wuestman M, Hoekman J and Frenken K (2020) A typology of scientific breakthroughs. *Quantitative Science Studies* 1(3): 1203–1222.